\documentclass[twocolumn, 11pt]{article}
\usepackage{array, float}
\usepackage[left=18mm, right=18mm, top=26mm, bottom=24mm]{geometry}
\usepackage{newtxtext}
\usepackage{amsmath, amssymb}
\usepackage{mathrsfs}
\usepackage{tabularx}
\usepackage{graphicx}
\usepackage{dcolumn}
\usepackage{setspace}
\usepackage{pgfpages}
\makeatletter
\usepackage{dcolumn}
\usepackage{url}
\usepackage{comment}
\usepackage{lineno}
\usepackage{cite}

\usepackage{abstract}

\usepackage{setspace}
\usepackage[font=small, format=plain, labelfont=bf, up, textfont=normal, up, justification=justified,singlelinecheck=false]{caption}

\usepackage[affil-it]{authblk}

\let\OLDthebibliography\thebibliography
\renewcommand\thebibliography[1]{
\OLDthebibliography{#1}
\setlength{\parskip}{0pt}
\setlength{\itemsep}{0pt plus 0.3ex}
}

\title{Relative opinion similarity leads to the emergence of large clusters in opinion formation models}
\author{Hirofumi Takesue\thanks{Electronic address: \texttt{hir.takesue@gmail.com}}}
\affil{Faculty of Law and Politics, Tokyo Metropolitan University}
\date{}

\begin{document}

\twocolumn[

\maketitle

\begin{onecolabstract}
This study considers a variant of the bounded confidence opinion formation model wherein the probability of opinion assimilation is dependent on the relative similarity of opinions. Agents are located on a social network and decide whether or not they adopt the opinion of one of the neighbors (called a role agent). Opinion assimilation is more (less) likely to occur when the distance from the opinion of the role agent is smaller (larger) than the average opinion distance from other neighbors. Thus, assimilation probability is reliant not only on opinion proximity with the role agent considered in conventional models but also on relative similarity that considers other neighbors. The simulation results demonstrate that large weights on relative similarity in determining assimilation probability increase the size of the largest opinion cluster. The size of the threshold parameter of the bounded confidence model displays inverse-U relationships with the largest cluster size. The findings imply that consideration of relative opinion similarity, as observed in recent empirical studies, prevents polarization into small opinion clusters.
\\\\
\end{onecolabstract}
]
\saythanks

\section*{Introduction}
Opinion formation is an intriguing collective phenomenon, where micro-level behavior is linked to macro-level patterns. Social attributes, including opinions, are formed through mutual social influence, such as conformity \cite{Cialdini2004, Izuma2013}. The accumulation of interactions leads to the emergence of various types of opinion distribution such as consensus and polarization. In this regard, opinion formation models specify assumptions about the micro-level process of social influence and examines the resultant macro-level opinion distribution \cite{Flache2017}. Researchers from various fields of physical, social, and computational sciences contribute to the understanding of the emergence process of opinion distribution \cite{Galam2008, Castellano2009, Flache2017}. The framework is also applied to topics of increasing importance such as the emergence of echo chambers \cite{Baumann2020, Brooks2020}.

One of the widely examined opinion formation models is the bounded confidence model \cite{Deffuant2000, Hegselmann2002}. In this model, each individual has a continuous variable called an opinion, and may assimilate their opinions through interactions with other agents. An important element of this model is a threshold parameter. In an interaction between two agents, assimilation occurs only when the opinion distance between the two agents is smaller than the exogenous threshold. Substantively, this assumption implies that extremely different opinions are not considered in opinion updating. Large (small) values of this threshold urge (hinder) the assimilation of opinions. This dynamic process leads to emergence of clusters of opinions, and the modification of opinions ceases once the opinion distance from agents outside the cluster becomes greater than the threshold. As a consequence, large threshold values generate small numbers of populous clusters, whereas small threshold values lead to polarized states with many small clusters \cite{Deffuant2000, Hegselmann2002, Weisbuch2002}. Realistic network structure is introduced to investigate its role in the opinion formation \cite{Fortunato2004, Stauffer2004, Meng2018, Schawe2021}.

An assumption shared by many opinion formation models is that the opinion distance between agents is an \textit{absolute} one (Ref. \cite{Mas2013} is a notable exception that considers relative similarity in modeling homophily). In pairwise interactions, the opinion distance between two interacting agents is calculated using the absolute difference in two opinion values. An implication of this assumption is that the opinions of other agents are not considered in calculating distance. Regardless of the opinions of others, the distance from the interacting agent is perceived to be the same. This fact can also play a role in the bounded confidence model, because the comparison between opinion distance and the threshold value determines whether or not opinion assimilations occur.

In contrast to this modeling assumption, recent empirical studies demonstrated that the perception of the position on (ideological) space is influenced by the opinion of a \textit{third-party}. A politically important example is the consideration of an extreme candidate. When an extreme alternative is considered within an ideological spectrum, other political actors are perceived to be more centrist \cite{Waismel-Manor2017, Wang2019e}. Moreover, extreme alternatives increase support for other moderate policies \cite{Simonovits2017}. The spatial voting model, which is a canonical one in economics and political science, assumes that people prefer closer alternatives in a political space, because the distance of opinions decreases utility \cite{Downs1957}. Following this logic, the observations suggest that the existence of other distant alternatives make people perceive the positions of other alternatives as more proximate.

Motivated by these findings, the current study considers \textit{relative similarity} in the bounded confidence model. In this model of opinion formation on a social network, the opinion of an agent (called a focal agent) may assimilate into an opinion of one of the neighbors (called a role agent). The classic models assume that whether or not opinion assimilation occurs is dependent on the opinion distance between the focal and role agents \cite{Flache2017}. In contrast, this study assumes that the occurrence of opinion assimilation is also dependent on the opinion distance between the focal agent and other neighbors on a network. The modified opinion distance becomes smaller (larger) when the distance from the opinion of the role agent is smaller (larger) than the average distance from the opinions of other neighbors. As a result, assimilation is more likely to occur when the focal and role agents share a relatively higher level of similarity than the focal agent and other neighbors. This mechanism potentially leads to the promotion and hindrance of opinion assimilation, because similar (distant) opinions tend to be perceived as more similar (distant).

Monte Carlo simulations demonstrate that the large influence of relative similarity on the probability of assimilation increased the size of the largest opinion cluster and generated less polarized opinion distribution. A scrutiny of the simulation process suggested that large weights on relative similarity foster the moderation of opinions, that is, assimilation to center positions. Given the specific value of opinion distance between the focal and role agents, assimilation into moderate opinion occurred with a higher probability than assimilation into extreme opinions. This bias toward moderation tendency helped the emergence of large clusters. These results demonstrated that the perception of opinion distance at the individual level can influence macroscopic opinion distribution. The literature proposed that small threshold values generate many small opinion clusters, but the current study suggests that this effect can be offset by a consideration of relative similarity and that opinion distribution does not easily become fragmented.

The study introduces heterogeneity to the opinion formation process, because assimilation probability can differ given the same threshold value and opinion distance from the role agent.  Previous studies also considered heterogeneity and demonstrated that the number of resultant clusters decreases when agents possess heterogeneous threshold levels \cite{Lorenz2010, Liang2013, Pineda2015, Han2019b}, although a few exceptions may exist \cite{Fu2015}. Relatedly, scholars also examined the effects of stubborn or external agents who (tend to) stick to their original opinions \cite{Weisbuch2005, McKeown2006, Pineda2015, Mathias2016}. The current study also assumed that assimilation possibility is heterogeneous even with constant opinion distance between paired agents. In contrast to the abovementioned studies, this research introduced dynamic heterogeneity generated by the opinion distribution of neighbors.

\section*{Model}
We considered a dynamic opinion formation process on small-world networks \cite{Watts1998}, which were constructed as follows \cite{Newman2000}. First, we generated an expanded cycle, where $N$ agents were connected with $z/2$ neighboring agents on both sides. Second, we added random $Nzp/2$ edges. The average degree (i.e., number of neighboring agents) is $z(1+p)$. Larger values of $p$ lead to the emergence of more disordered networks. Small-world networks facilitate consensus in various opinion formation models that assume opinion assimilation \cite{Flache2011, Takesue2021a}.

Each agent ($i$) possesses a trait called an opinion, which is denoted as $o_i$. Opinions represent a position on a (political) spectrum such as ideology and policy attitudes. For instance, large (small) opinion values can correspond to right (left) positions on an ideological spectrum. The initial values of opinions are sampled from the standard uniform distribution: $\mathrm{U}(0, 1)$.

Agents can update their opinions through the interaction process inspired by the bounded confidence model. For each round, one agent (the focal agent) is randomly selected from the entire population, and one of the neighbors (the role agent) is also randomly selected. The assimilation of the opinion of the focal agent occurs with the following probability:
\begin{equation*}
p_{o_f \leftarrow o_r} = 1/[1 + \exp(\beta (\hat{d}_{fr} - \delta))],
\end{equation*}
where $\hat{d}_{fr}$ is the modified opinion distance, which will be explained below, and $\delta$ is a threshold parameter. This setting assumes that interactions with a more distant neighbor decrease assimilation probability, but this distance is tolerated by $\delta$. In extreme cases, where $\beta \to \infty$, the threshold parameter $\delta$ plays the same role as that in the conventional bounded confidence model, because assimilation probability changes from 1 to 0 when $\hat{d}_{ij}$ exceeds $\delta$. Scholars frequently adopt this type of functional form to represent a realistic noisy process of opinion adoption \cite{Grauwin2012}.

The modified opinion distance consists of two types of distance. The first is the distance between the opinion of the focal agent ($o_f$) and that of the role agent ($o_r$). The opinion distance between the two agents is denoted as $d_{fr}$. The second one is average opinion distance between the focal agent and her neighbors (excluding the role agent). It is calculated as $d_{f\mathcal{N}_f} = \sum_{k \in \mathcal{N}_f\setminus r} |o_f - o_k|/(z_f - 1)$, where summation runs over the neighborhood of the focal agents ($\mathcal{N}_f$), excluding the role agent, and $z_f$ is the degree of the focal agent. Based on these values, the modified distance is calculated as follows:
\begin{equation*}
\hat{d}_{fr} = d_{fr} + \alpha (d_{fr} - d_{f\mathcal{N}_f}).
\end{equation*}
The modified distance decreases (and assimilation probability increases) when the opinion distance between the focal and role agents is smaller than the average distance between the focal agent and her neighbors. At this point, we introduce relative similarity in the model. The parameter, $\alpha (\geq 0)$, is a weight on this relative similarity, and distance becomes a conventional one when $\alpha = 0$. The primary interest of the study is in examining the role of this parameter.

If opinion assimilation occurs, then the opinion of the focal agent approaches that of the role agent. Specifically, opinion updating occurs as follows:
\begin{equation*}
o_f \leftarrow o_f + \mu (o_r - o_f),
\end{equation*}
where $\mu$ is the assimilation intensity. Scholars frequently assume that the opinions of the two agents become the same value after interaction \cite{Deffuant2000}; therefore, the current study assumes that $\mu = 1$.

The focal agent adopts a random opinion with a small probability ($p_e$) instead of interacting with others. In this case, a new value of $o_f$ is sampled from $\mathrm{U}(0, 1)$. Previous studies demonstrate that the polarized state with multiple opinion clusters is fragile in the bounded confidence model (and other opinion formation models) if agents are allowed to interact with individuals outside the cluster with a (small) probability. Entirely homogeneous states emerge even with a small threshold value in this case \cite{Klemm2003a, Kurahashi-Nakamura2016}. However, the original pattern is restored if individuals adopt a random opinion with a small probability \cite{Kurahashi-Nakamura2016}. This noise prevents the emergence of entirely homogeneous stats while allowing realistic noisy opinion formation processes.

We examined the behavior of this model using Monte Carlo simulation. The main quantity of interest is the size of the largest opinion cluster divided by $N$, which is denoted as $S_{\max}$. This parameter is frequently adopted as the order parameter \cite{Lorenz2010, Liang2013, Pineda2015}. The relaxing process continued for at least 50 000$N$ rounds. Moreover, the sampling process continued for at least 300 000$N$ rounds to achieve statistical accuracy. We conducted at least 10 simulation runs for each combination of parameter values and reported the mean values of these runs.

\section*{Results}
\begin{figure}[tbp]
\centering
\vspace{5mm}
\includegraphics[width = 65mm, trim= 20 20 0 0]{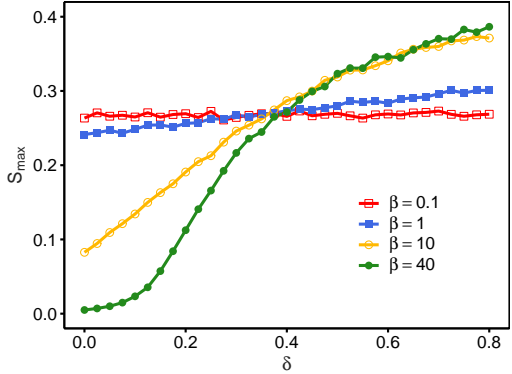}
\caption{\small The figure reports the effect of the threshold parameter ($\delta$) when the weight on relative similarity is absent ($\alpha = 0$). Large $\delta$ values give rise to large opinion clusters. The other parameters are $N = 5000, z = 4, p = 0.02$, and $p_e = 10^{-4}$.}
\label{fig_S_d_beta}
\end{figure}

The first analysis reported in Figure \ref{fig_S_d_beta} examined the behavior of the model without relative similarity (i.e., $\alpha = 0$). The threshold parameter ($\delta$) illustrates the expected effects on $S_{\max}$ in the same manner as the orthodox bounded confidence model despite differences in the detailed model specification. Large $\delta$ values permit opinion assimilation with distant individuals and contributes to the emergence of large opinion clusters. This pattern is prominent with a small noise (large $\beta$). The decreasing effects of $\delta$ is observed, and the values of $S_{\max}$ fails to reach one due to the random adoption of opinions. However, the basic role of $\delta$ remains similar. The figure suggests that $\beta = 10$ is sufficiently large to confirm the effect of $\delta$. Thus, we set the value of $\beta$ to 10 in the following analysis.

\begin{figure}[tbp]
\centering
\vspace{5mm}
\includegraphics[width = 65mm, trim= 20 20 0 0]{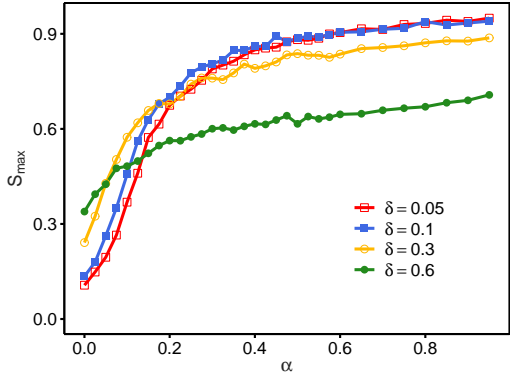}
\caption{\small The figure notes the effect of the weight on relative similarity ($\alpha$). The size of the largest cluster increases with large $\alpha$ values. Moreover, this effect is more prominent with small threshold values ($\delta$). Other parameters include $N = 5000, z = 4, p = 0.02, \beta = 10$, and $p_e = 10^{-4}$.}
\label{fig_S_alpha_d}
\end{figure}

Figure \ref{fig_S_alpha_d} depicts that the large weight on relative similarity leads to large opinion clusters. This figure reports the largest cluster size as a function of the main parameter, namely, weight on relative similarity ($\alpha$), for different threshold values ($\delta$). Large $\alpha$ values increase the size of the largest cluster. We examined the cases with larger $\alpha$ values that are not reported in the figure and confirmed that effect of $\alpha$ is monotonically positive (despite decreasing effect size) at least when $\alpha \leq 2$. The figure also illustrates that the positive effect of $\alpha$ is more prominent with a small threshold ($\delta$).

\begin{figure}[tbp]
\centering
\vspace{5mm}
\includegraphics[width = 65mm, trim= 20 20 0 0]{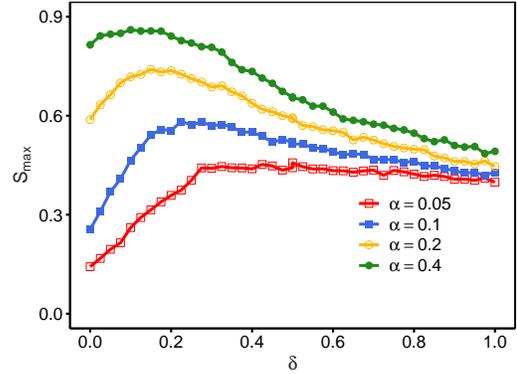}
\caption{\small The figure points to the effect of the threshold parameter ($\delta$) and frequently observes non-monotonic patterns in contrast to the case of $\alpha = 0$. The other parameters are $N = 5000, z = 4, p = 0.02, \beta = 10$, and $p_e = 10^{-4}$.}
\label{fig_S_d_alpha}
\end{figure}

Figure~\ref{fig_S_d_alpha} reports the relationships between $S_{\max}$ and the threshold parameter ($\delta$). In contrast to the pattern in Figure \ref{fig_S_d_beta} and the conventional model, we did not necessarily observe the monotonic positive effect of $\delta$. As the value of $\alpha$ is sufficiently large (0.1 in Figure ~\ref{fig_S_d_alpha}), $S_{\max}$ frequently decreases with the increase in $\delta$. We observed the small effects of $\alpha$ as a result of this decreasing trend. With further large $\delta$ values that are not reported in the figure (e.g., $\delta = 2$), the effects of $\alpha$ become nearly negligible. Large tolerance levels solely determine the assimilation probability, and the contribution of $\alpha$ (and modified opinion distance) to assimilation probability decreases.

The next analysis focuses on the underlying mechanism of the pattern, that is, the weight on relative similarity exerts positive effects on the largest cluster size. Specifically, we focus on the opinion moderation tendency with positive $\alpha$ values. We call the scenario opinion moderation (extremization), when an opinion moves closer to (away from) the center position (i.e., 0.5). If the opinion position of the role agent is closer to (far from) the center than the original position of the focal agent, then it is the \textit{opportunity} of moderation (extremization). We calculated the probability of opinion assimilation by dividing the number of \textit{realized} opinion moderation (extremization) by that of moderation (extremization) opportunities. We call this probability of opinion moderation (extremization) as $p_{\mathrm{mod}}$ ($p_{\mathrm{ext}}$).

\begin{figure}[tbp]
\centering
\vspace{5mm}
\includegraphics[width = 85mm, trim= 20 20 0 0]{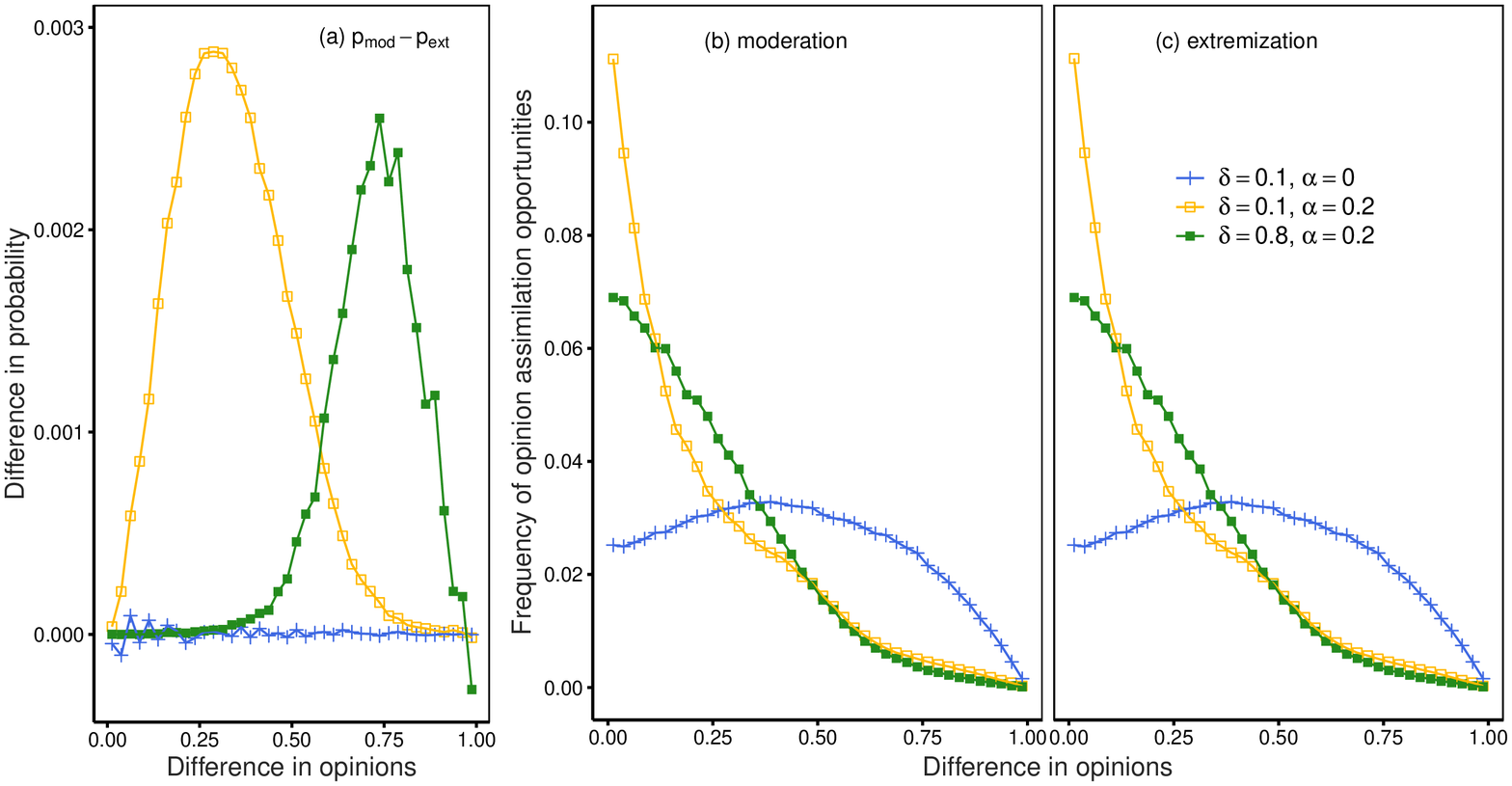}
\caption{\small Panel (a) reports the differences between moderation and extremization probability as a function of opinion distance between the focal and role agents. Opinion moderation occurs with a high probability when $\alpha > 0$. Panels (b) and (c) report the frequencies of assimilation opportunities. The study conducted 1500 (500) simulation runs for the cases of $\delta = 0.8$ and $\alpha = 0.2$ (other parameter values). Each run continued for 20 000 rounds. We recorded the number of opportunities and realizations of opinion assimilation by the values of opinion distance between the two agents. Opinion distance was grouped in the range of 0.025. The other parameters include $N = 5000, z = 4, p = 0.02, \beta = 10$ and $p_e = 10^{-4}$.}
\label{fig_direc_prob}
\end{figure}

Figure~\ref{fig_direc_prob} reports bias toward opinion moderation with positive $\alpha$. Panel (a) of the figure reports $p_{\mathrm{mod}} - p_{\mathrm{ext}}$ as a function of the absolute opinion difference between the focal and role agents. The case of $\alpha = 0$ illustrates that the moderation and extremization of opinions occurred without bias. In contrast, opinion moderation was more likely to occur given a specific opinion distance when $\alpha > 0$. This fact could contribute to the emergence of a large opinion cluster through the accumulation of opinions near the center position.

This bias may also be related to the attenuated effects of $\alpha$ with large $\delta$. A comparison of cases with the same $\alpha$ value (0.2) but different $\delta$ values (0.1 and 0.8) in Panel (a) indicates that bias takes the maximum values with different values of opinion distance. Moreover, this factor could influence cluster size through different frequencies of opinion modifications. Panels (b) and (c) report the frequency of moderation and extremization opportunities. In the cases of positive $\alpha$, assimilation opportunities occur with high frequencies when the focal and role agents share similar opinions. This pattern is observed regardless of the direction of movement. Bias toward moderation is observed with relatively large differences in opinion in the case of $\delta = 0.8$ (Panel (a)), but less assimilation opportunities are observed between distant opinions (panels (b) and (c)). As a result, the effects of moderation tendency become weak with a large threshold, which may explain the small $S_{\max}$ with large $\delta$.

This bias toward moderation can be heuristically understood using a simple example of initial states, where opinions follow a uniform distribution. The average opinion distance from other agents takes the minimum values at the center position and increases as the opinion moves away from the center. Therefore, agents near the center tend to possess relatively similar opinions when they become role agents. The shape of opinion distribution changes through the dynamic process, and this inference is not directly applicable to this process. However, the advantage of the center position can persist given that opinions are not skewed on both ends.

\begin{figure}[tbp]
\centering
\vspace{5mm}
\includegraphics[width = 85mm, trim= 20 20 0 0]{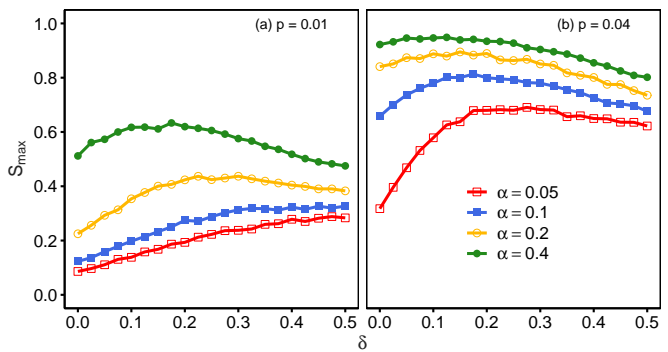}
\caption{\small The largest cluster size is reported with different values of $p$. The positive effects of $\alpha$ and the non-monotonic effects of $\delta$ are replicated. More disordered networks (larger $p$) contribute to larger $S_{\max}$. The other parameters are $N = 5000, z = 4, \beta = 10$, and $p_e = 10^{-4}$.}
\label{fig_S_d_alpha_p}
\end{figure}

Finally, we changed the values of the other parameters and examined the robustness of the reported patterns. Figure \ref{fig_S_d_alpha_p} reports results with different values of $p$, which controls the number of added links in generating small-world networks. A comparison of the panels indicates that large $p$ (more disordered networks) tend to give rise to larger clusters. This observation is consistent with the findings that disordered networks are more likely to reach consensus \cite{Flache2011, Takesue2021a}. Importantly, the figure replicates the basic reported patterns, that is, the weight on relative similarity ($\alpha$) increases $S_{\max}$, whereas the tolerance parameter ($\delta$) exhibits non-monotonic effects.

\begin{figure}[tbp]
\centering
\vspace{5mm}
\includegraphics[width = 85mm, trim= 20 20 0 0]{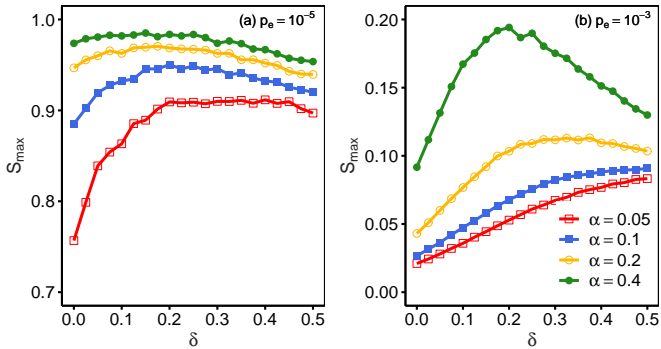}
\caption{\small The largest cluster size is reported with different values of $p_e$. The positive effects of $\alpha$ and the non-monotonic effects of $\delta$ are replicated. The small probability of adopting random opinion values increases $S_{\max}$. Other parameters include $N = 5000, z = 4, p = 0.02$, and $\beta = 10$.}
\label{fig_S_d_alpha_pe}
\end{figure}

We also conducted simulations with different values of $p_e$, that is, the probability of adopting random opinions (Figure \ref{fig_S_d_alpha_pe}). The overall patterns suggested that a decrease in $p_e$ leads to larger $S_{\max}$ (notably, the scale of y-axis differs between the two panels). This result is natural, because the errors dismantle the existing clusters. In the special cases of $p_e = 0$, the system always converged to entirely homogeneous opinion distributions, because assimilation between distant opinions can occur with a positive probability \cite{Kurahashi-Nakamura2016}. We also confirmed the results reported in this work: the positive effects of $\alpha$ and non-monotonic effects of $\delta$.

\section*{Conclusion}
The study examined the role of relative similarity in an opinion formation model. In the model, probability of opinion assimilation is dependent not only on the opinion of a role agent, whose opinion may be imitated, but also the opinions of other neighbors in the networks. Specifically, assimilation probability increases (decreases) when the opinion distance from the role agent is smaller (larger) than the average distance from other neighbors. As a result, opinion assimilation of the opinion of the role agent is more likely to occur when she has a relatively similar opinion. The simulations demonstrated that larger weights on this relative similarity give rise to the increase in the size of the largest cluster. The threshold parameter in the bounded confidence model depicted non-monotonic effects wherein the size of the largest cluster reaches its maximum value with a moderate value of the threshold parameter. These patterns can be explained by the tendency toward opinion moderation with positive weights on relative similarity. Given the opinion distance from the role agent, opinion assimilation toward the center position occurs with a higher probability than that toward the extreme positions.

The proposed model suggests that opinion distribution is not easily fragmented once the opinion of neighbors is considered in determining distance. The underlying mechanism of this observation is consistent with those of recent empirical studies. A recent study finds that the existence of other extreme alternatives increases the support for other policies that lie at the (relatively) centrist position \cite{Simonovits2017}. In the simulation process, positive weights on relative opinion similarity generate the moderation tendency of opinion assimilation. Although the model did not explicitly assume the popularity of the centrist opinion, a consideration of relative similarity renders centrist opinions more attractive.

Lastly, we discuss the limitations and potential extension of this study. The model is not the sole method for considering relative similarity, such that different settings should be considered. This study applied relative similarity in calculating opinion assimilation probability. However, a possibility existed that the magnitude of opinion change is also affected. In addition, the robustness of the results must be considered when combined with other factors. In this model, the emergence of the large cluster was fostered by opinion moderation. Other opinion formation models introduced repulsive opinion modification \cite{Jager2005, VazMartins2010, Chen2017d, Han2022b}. Relatedly, stubborn agents \cite{Weisbuch2005, Galam2007, Sobkowicz2015, Mathias2016} and multiple opposing mass media \cite{McKeown2006, Takesue2021a} can induce polarization. Thus, the role of relative similarity can be examined when combined with the mechanism that acts in the opposite direction. Furthermore, the idea of relative similarity can be applied to other opinion formation models given that opinion distance is defined. Candidates are continuous opinions \cite{Fortunato2005a, Baldassarri2007}, ordered discrete opinions \cite{Ben-Naim2003, Stauffer2004a}, and opinions consisting of a vector of (unordered) attributes \cite{Axelrod1997}. The effects of relative similarity may depend on the types of opinion models. In this regard, the consideration of relative similarity is an empirically observed phenomenon, and further examining the influence of the this factor on opinion formation processes may be meaningful.


\begin{thebibliography}{00}

\bibitem{Cialdini2004}
Robert~B. Cialdini and Noah~J. Goldstein.
\newblock {Social influence: Compliance and conformity}.
\newblock {\em Annual Review of Psychology}, 55(1974):591--621, 2004.

\bibitem{Izuma2013}
Keise Izuma.
\newblock {The neural basis of social influence and attitude change.}
\newblock {\em Current opinion in neurobiology}, 23(3):456--62, 2013.

\bibitem{Flache2017}
Andreas Flache, Michael M{\"{a}}s, Thomas Feliciani, Edmund Chattoe-Brown,
  Guillaume Deffuant, Sylvie Huet, and Jan Lorenz.
\newblock {Models of Social Influence: Towards the Next Frontiers}.
\newblock {\em Journal of Artificial Societies and Social Simulation},
  20(4):3521, 2017.

\bibitem{Galam2008}
Serge Galam.
\newblock {Sociophysics: A review of galam models}.
\newblock {\em International Journal of Modern Physics C}, 19(3):409--440,
  2008.

\bibitem{Castellano2009}
Claudio Castellano, Santo Fortunato, and Vittorio Loreto.
\newblock {Statistical physics of social dynamics}.
\newblock {\em Reviews of Modern Physics}, 81(2):591--646, 2009.

\bibitem{Baumann2020}
Fabian Baumann, Philipp Lorenz-Spreen, Igor~M. Sokolov, and Michele Starnini.
\newblock {Modeling Echo Chambers and Polarization Dynamics in Social
  Networks}.
\newblock {\em Physical Review Letters}, 124(4):48301, 2020.

\bibitem{Brooks2020}
Heather~Z. Brooks and Mason~A. Porter.
\newblock {A model for the influence of media on the ideology of content in
  online social networks}.
\newblock {\em Physical Review Research}, 2(2):023041, 2020.

\bibitem{Deffuant2000}
Guillaume Deffuant, David Neau, Frederic Amblard, and G{\'{e}}rard Weisbuch.
\newblock {Mixing beliefs among interacting agents}.
\newblock {\em Advances in Complex Systems}, 03(01n04):87--98, 2000.

\bibitem{Hegselmann2002}
Rainer Hegselmann and Ulrich Krause.
\newblock {Opinion Dynamics and Bounded Confidence}.
\newblock {\em Journal of Artificial Societies and Social Simulation}, 5(3):2,
  2002.

\bibitem{Weisbuch2002}
G{\'{e}}rard Weisbuch, Guillaume Deffuant, Fr{\'{e}}d{\'{e}}ric Amblard, and
  Jean-Pierre Nadal.
\newblock {Meet, discuss, and segregate!}
\newblock {\em Complexity}, 7(3):55--63, 2002.

\bibitem{Fortunato2004}
Santo Fortunato.
\newblock {Universality of the Threshold for Complete Consensus for the Opinion
  Dynamics of Deffuant et al.}
\newblock {\em International Journal of Modern Physics C}, 15(09):1301--1307,
  2004.

\bibitem{Stauffer2004}
D.~Stauffer and H.~Meyer-Ortmanns.
\newblock {Simulaton of consensus model of Deffuant et al. on a
  Barab{\'{a}}si-Albert network}.
\newblock {\em International Journal of Modern Physics C}, 15(02):241--246,
  2004.

\bibitem{Meng2018}
X~Flora Meng, Robert~A. {Van Gorder}, and Mason~A Porter.
\newblock {Opinion formation and distribution in a bounded-confidence model on
  various networks}.
\newblock {\em Physical Review E}, 97(2):022312, 2018.

\bibitem{Schawe2021}
Hendrik Schawe, Sylvain Fontaine, and Laura Hern{\'{a}}ndez.
\newblock {When network bridges foster consensus. Bounded confidence models in
  networked societies}.
\newblock {\em Physical Review Research}, 3(2):023208, 2021.

\bibitem{Mas2013}
Michael M{\"{a}}s and Andreas Flache.
\newblock {Differentiation without Distancing. Explaining Bi-Polarization of
  Opinions without Negative Influence}.
\newblock {\em PLoS ONE}, 8(11):e74516, 2013.

\bibitem{Waismel-Manor2017}
Israel Waismel-Manor and Gabor Simonovits.
\newblock {The Interdependence of Perceived Ideological Positions}.
\newblock {\em Public Opinion Quarterly}, 81(3):759--768, 2017.

\bibitem{Wang2019e}
Austin Horng~En Wang and Fang~Yu Chen.
\newblock {Extreme Candidates as the Beneficent Spoiler? Range Effect in the
  Plurality Voting System}.
\newblock {\em Political Research Quarterly}, 72(2):278--292, 2019.

\bibitem{Simonovits2017}
Gabor Simonovits.
\newblock {Centrist by Comparison: Extremism and the Expansion of the Political
  Spectrum}.
\newblock {\em Political Behavior}, 39(1):157--175, 2017.

\bibitem{Downs1957}
Anthony Downs.
\newblock {\em {An Economic Theory of dDemocracy}}.
\newblock Harper {\&} Row, New York, 1957.

\bibitem{Lorenz2010}
Jan Lorenz.
\newblock {Heterogeneous bounds of confidence: Meet, discuss and find
  consensus!}
\newblock {\em Complexity}, 15(4):43--52, 2010.

\bibitem{Liang2013}
Haili Liang, Yiping Yang, and Xiaofan Wang.
\newblock {Opinion dynamics in networks with heterogeneous confidence and
  influence}.
\newblock {\em Physica A: Statistical Mechanics and its Applications},
  392(9):2248--2256, 2013.

\bibitem{Pineda2015}
M.~Pineda and G.~M. Buend{\'{i}}a.
\newblock {Mass media and heterogeneous bounds of confidence in continuous
  opinion dynamics}.
\newblock {\em Physica A: Statistical Mechanics and its Applications},
  420:73--84, 2015.

\bibitem{Han2019b}
Wenchen Han, Changwei Huang, and Junzhong Yang.
\newblock {Opinion clusters in a modified Hegselmann-Krause model with
  heterogeneous bounded confidences and stubbornness}.
\newblock {\em Physica A: Statistical Mechanics and its Applications},
  531:121791, 2019.

\bibitem{Fu2015}
Guiyuan Fu, Weidong Zhang, and Zhijun Li.
\newblock {Opinion dynamics of modified Hegselmann-Krause model in a
  group-based population with heterogeneous bounded confidence}.
\newblock {\em Physica A: Statistical Mechanics and its Applications},
  419:558--565, 2015.

\bibitem{Weisbuch2005}
G{\'{e}}rard Weisbuch, Guillaume Deffuant, and Fr{\'{e}}d{\'{e}}ric Amblard.
\newblock {Persuasion dynamics}.
\newblock {\em Physica A: Statistical Mechanics and its Applications},
  353(1-4):555--575, 2005.

\bibitem{McKeown2006}
Gary Mckeown and Noel Sheehy.
\newblock {Mass media and polarisation processes in the bounded confidence
  model of opinion dynamics}.
\newblock {\em Journal of Artificial Societies and Social Simulation},
  9(1):33--63, 2006.

\bibitem{Mathias2016}
Jean-Denis Mathias, Sylvie Huet, and Guillaume Deffuant.
\newblock {Bounded Confidence Model with Fixed Uncertainties and Extremists:
  The Opinions Can Keep Fluctuating Indefinitely}.
\newblock {\em Journal of Artificial Societies and Social Simulation}, 19(1):6,
  2016.

\bibitem{Watts1998}
Duncan~J Watts and Steven~H Strogatz.
\newblock {Collective dynamics of `small-world' networks}.
\newblock {\em Nature}, 393(6684):440--442, 1998.

\bibitem{Newman2000}
M.~E.~J. Newman, C.~Moore, and D.~J. Watts.
\newblock {Mean-field solution of the small-world network model}.
\newblock {\em Physical Review Letters}, 84(14):3201--3204, 2000.

\bibitem{Flache2011}
Andreas Flache and Michael~W. Macy.
\newblock {Small Worlds and Cultural Polarization}.
\newblock {\em The Journal of Mathematical Sociology}, 35:146--176, 2011.

\bibitem{Takesue2021a}
Hirofumi Takesue.
\newblock {A Noisy Opinion Formation Model with Two Opposing Mass Media}.
\newblock {\em Journal of Artificial Societies and Social Simulation}, 24(4):3,
  2021.

\bibitem{Grauwin2012}
S{\'{e}}bastian Grauwin and Pablo Jensen.
\newblock {Opinion group formation and dynamics: Structures that last from
  nonlasting entities}.
\newblock {\em Physical Review E}, 85(6):066113, 2012.

\bibitem{Klemm2003a}
Konstantin Klemm, V{\'{i}}ctor~M. Egu{\'{i}}luz, Ra{\'{u}}l Toral, and Maxi~San
  Miguel.
\newblock {Global culture: A noise-induced transition in finite systems}.
\newblock {\em Physical Review E}, 67(4):045101, 2003.

\bibitem{Kurahashi-Nakamura2016}
Takasumi Kurahashi-Nakamura, Michael M{\"{a}}s, and Jan Lorenz.
\newblock {Robust Clustering in Generalized Bounded Confidence Models}.
\newblock {\em Journal of Artificial Societies and Social Simulation}, 19(4),
  2016.

\bibitem{Jager2005}
Wander Jager and Fr{\'{e}}d{\'{e}}ric Amblard.
\newblock {Uniformity, bipolarization and pluriformity captured as generic
  stylized behavior with an agent-based simulation model of attitude change}.
\newblock {\em Computational {\&} Mathematical Organization Theory},
  10(4):295--303, 2005.

\bibitem{VazMartins2010}
T.~{Vaz Martins}, Miguel Pineda, and Raul Toral.
\newblock {Mass media and repulsive interactions in continuous-opinion
  dynamics}.
\newblock {\em EPL (Europhysics Letters)}, 91(4):48003, 2010.

\bibitem{Chen2017d}
Guodong Chen, Hongyan Cheng, Changwei Huang, Wenchen Han, Qionglin Dai, Haihong
  Li, and Junzhong Yang.
\newblock {Deffuant model on a ring with repelling mechanism and circular
  opinions}.
\newblock {\em Physical Review E}, 95(4):042118, 2017.

\bibitem{Han2022b}
Wenchen Han, Shun Gao, Changwei Huang, and Junzhong Yang.
\newblock {Non-consensus states in circular opinion model with repulsive
  interaction}.
\newblock {\em Physica A: Statistical Mechanics and its Applications},
  585:126428, 2022.

\bibitem{Galam2007}
Serge Galam and Frans Jacobs.
\newblock {The role of inflexible minorities in the breaking of democratic
  opinion dynamics}.
\newblock {\em Physica A: Statistical Mechanics and its Applications},
  381(1-2):366--376, 2007.

\bibitem{Sobkowicz2015}
Pawel Sobkowicz.
\newblock {Extremism without extremists: Deffuant model with emotions}.
\newblock {\em Frontiers in Physics}, 3(March):17, 2015.

\bibitem{Fortunato2005a}
Santo Fortunato.
\newblock {The Sznajd consensus model with continuous opinions}.
\newblock {\em International Journal of Modern Physics C}, 16(1):17--24, 2005.

\bibitem{Baldassarri2007}
Delia Baldassarri and Peter Bearman.
\newblock {Dynamics of Political Polarization}.
\newblock {\em American Sociological Review}, 72(5):784--811, 2007.

\bibitem{Ben-Naim2003}
E.~Ben-Naim, P.L. Krapivsky, and S~Redner.
\newblock {Bifurcations and patterns in compromise processes}.
\newblock {\em Physica D: Nonlinear Phenomena}, 183(3-4):190--204, 2003.

\bibitem{Stauffer2004a}
Dietrich Stauffer, Adriano Souza, and Christian Schulze.
\newblock {Discretized Opinion Dynamics of The Deffuant Model on Scale-Free
  Networks}.
\newblock {\em Journal of Artificial Societies and Social Simulation}, 7(3):7,
  2004.

\bibitem{Axelrod1997}
Robert Axelrod.
\newblock {The Dissemination of Culture: A Model with Local Convergence and
  Global Polarization}.
\newblock {\em Journal of Conflict Resolution}, 41(2):203--226, 1997.

\end{thebibliography}

\end{document}